\documentstyle[12pt]{article}
\begin{document}

\begin{titlepage}

\begin{flushright}
INFN-FE 12-96 \\ 
September 1996 
\end{flushright}

\vspace{1cm}

\begin{center}
{\Large\bf SUSY SU(6) GUT \\
without Gauge Hierarchy Problem}

\vspace{3mm}

{\bf Zurab \ Tavartkiladze \footnote {E-mail address:
tavzur@axpfe1.fe.infn.it}
}
\vspace{3mm}

{\it INFN Sezione di Ferrara, I-44100 Ferrara, Italy \\
and Institute of Physics of Georgian Academy of Sciences,
380077 Tbilisi, Georgia
}
\end{center}

\begin{abstract} 
A solution of the doublet-triplet splitting problem in the supersymmetric
$SU(6)$ gauge theory is suggested. The `missing doublet' multiplet
-- 175-plet of the $SU(6)$ group as well as the custodial 
$SU(2)_{cus}$ global
symmetry play crucial role for achieving the doublet-triplet hierarchy.
Two examples in which the doublet-triplet splitting occurs 
naturally are presented.
\end{abstract}

\end{titlepage}

The supersymmetric (SUSY) Grand Unified 
Theories (GUT) provide an attractive possibility to
understand the stability of the electroweak symmetry breaking
scale and the unification of the gauge couplings.
It is well known \cite{lan} that in
the minimal supersymmetric standard model (MSSM)
the constants $g_{3,2,1}$ of the gauge group 
$G_{321}\equiv SU(3)_C\times SU(2)_W\times U(1)_Y$ join at
energies $M_{X}\sim 10^{16}$ GeV, at which scale $G_{321}$
can be consistently embedded in $SU(5)$ or some larger group $G$.
This suggests the following paradigm: below the Planck scale
$M_{Pl}$ the hypothetical ``theory of everything'' reduces to a
SUSY GUT with gauge group $G$, which  first breaks 
down to $SU(5)$ at scale $M_G\geq M_X$, and then 
at the scale $M_X$ $SU(5)$
reduces to MSSM :
\begin{equation}
G \stackrel{M_G}{\rightarrow} SU(5) 
\stackrel{M_X}{\rightarrow} G_{321}          \label {pat}
\end{equation}
Obviously, it is also possible that the $G$ breaks to $G_{321}$
at once, directly at the scale $M_G\sim 10^{16}$ GeV.

The main problem which emerges in SUSY GUTs is a problem
of the doublet-triplet (DT) splitting.
The MSSM Higgs doublets ($h_{1,2}$) which induce
the electroweak symmetry breaking
and fermion masses should be light (with mass $\sim M_W$),
while their colour-triplet partners in GUT supermultiplets should 
have masses of order of $M_X$ in order to avoid too fast decay of nucleon.
%kodi 96407
Several mechanisms are known for solving the DT splitting
problem without fine tuning:

(i) The {\it missing partner} mechanism \cite{dim},
which is operative directly in $SU(5)$ theory. Besides the 
standard $\bar 5+5$ Higgses it requires the `missing doublet'
multiplets $\overline {50}+50$ (which however contain the 
colour-triplets) and the Higgs 75-plet for the $SU(5)$ breaking.

(ii) The {\it missing VEV} mechanism \cite{dim1} can be realized in 
$SO(10)$ model. Among other relevant Higgs multiplets it employs
two 10-plets and a 45-plet with the specific direction of
VEV towards the $B-L$ generator of $SO(10)$.

(iii) The {\it Goldstone boson} mechanism \cite{ino, ber}. 
In these scenarios the light Higgs doublets $h_{1,2}$ emerge as 
pseudo-goldstone modes, as a result of the spontaneous breaking 
of the  larger global symmetry of the superpotential.  
In particular, the models \cite{ber} based on $SU(6)$ gauge
symmetry includes Higgses 35 and $\bar 6+6$, and the Higgs
superpotential possess an accidental global symmetry $SU(6)\times SU(6)$   
independently transforming these two sets.

(iv) The {\it custodial symmetry} mechanism \cite{gia} is
also based on the $SU(6)$ gauge group.
The Higgs sector includes 
the 35-plet and two pairs of $\bar 6+6$
related by the 'custodial'  global symmetry $SU(2)_{cus}$.
If the mass term of 35 is suppressed in 
the superpotential (or it is the SUSY breaking scale 
$\sim M_W$),
then after $SU(6)$ breaking to $G_{321}$ the pair of
doublet-antidoublet from $\bar6+6$ which can serve as MSSM 
Higgses $h_{1,2}$ remain light.

In the case (i) the $SU(5)$ unification of the gauge constants is 
straightforward, with the possible uncertainties related to the 
GUT threshold corrections.  
In the cases (ii) and (iii) with $G=SO(10)$ and $SU(6)$
respectively the hierarchy $M_G\geq M_X$ in breaking (\ref {pat})
is consistent and even can have interesting understanding
of fermion mass hierarchies \cite{And, bar}. 
However, in the  custodial symmetry
 mechanism \cite{gia} the picture (\ref {pat}) can not be
achieved and the $SU(6)$ gauge group breaking proceeds as 
\begin{equation}
SU(6) \stackrel{M_G}{\rightarrow} SU(3)_C\times SU(3)_W\times U(1)_I 
\stackrel{M_I}{\rightarrow} G_{321}        \label {cus}
\end{equation}
where due to specifics of the model the intermediate scale
$M_I$ emerges as a geometrical 
$M_I\sim \sqrt {M_GM_W}$.
Consequently, unification of the gauge couplings is spoiled.

A possibility of  improving this drawback was suggested in \cite{175},
where the 35-plet of the model \cite{gia} 
was replaced by the 175-plet of $SU(6)$ --- the traceless tensor 
$\Phi^{ABC}_{A'B'C'}$ antisymmetric in the up and down indices. 
It is crucial that 175, in contrast to 35, does not contain the 
$G_{321}$ doublet fragments.
This feature allows to have $M_I\sim M_G$ and in principle the gauge
couplings could be directly unified at the scale $M_G\sim 10^{16}$ GeV.
However, there emerges the following problem: 
$\Phi$ has no renormalizable coupling to the Higgses $\bar H, H$ 
in representations $\bar 6+6$, so that the renormalizable Higgs 
superpotential possesses an extra global symmetry, related to the 
independent $SU(6)$ transformations of $\Phi$ 
and $\bar H , H$. In order to avoid the extra Goldstone degrees 
of freedom which in fact are the colour triplets, 
the nonrenormalizable couplings like 
$\frac {1}{M^2_P}\Phi^3H\bar H$ cutoff by the (reduced or genuine) Planck 
scale, $M_{P}\sim 10^{18}-10^{19}$ GeV, should be introduced in the theory. 
However, since $M_G\ll M_P$, these colour triplets will get the masses 
no more than $\sim M_G^3/M_{P}^2= 10^{10}-10^{12}$ GeV.
This would affect the unification of the gauge couplings
and, which is more dramatic, would lead to the unacceptably fast
proton decay.

In this paper we present the possibility which do not suffer from 
these problems. It naturally provides the DT splitting while 
the extra global symmetries can be avoided already 
at level of the renormalizable superpotential.
\vspace{2mm}

Consider the supersymmetric gauge $SU(6)$ model with the global 
custodial symmetry $SU(2)_{cus}$. 
The fermion sector  consists of the anomaly 
free chiral set of supermultiplets with the following content
under the $SU(6)\times SU(2)_{cus}$ group per one generation:  
$\bar 6^m \sim (\bar 6, \bar 2)$ and $15\sim (15, 1)$,  
where $m$ is the $SU(2)_{cus}$ index.

The Higgs sector contains the superfields $\Phi \sim (175,1)$ 
needed for breaking of $SU(6)$ to 
$G_{331}=SU(3)_C\times SU(3)_W\times U(1)_I$, and 
$\Psi_m\sim (R,2)$ and $\overline{\Psi}^m\sim (\bar{R},2)$, $m=1,2$, 
for the further breaking of $G_{331}$ down to $G_{321}$.
Here $R$ can be 84 or 210 of $SU(6)$, which representations 
are uniquely selected by the following requirements: 

\begin{itemize}
\item[a)] It should acquire a VEV inducing the $G_{331}$ symmetry 
breaking down to $G_{321}$. Therefore, it should contain the $G_{321}$ 
doublet fragment which will be absorbed by corresponding vector 
superfields of $G_{331}$ through the Higgs mechanism.
\item[b)] The tensor product $\overline {R}\times R $  should contain
175 in order to avoid the accidental global symmetries in the Higgs 
superpotential 
due to the renormalizable coupling $\Phi\Psi\overline{\Psi}$.  
\end{itemize}
 
Therefore, the most general renormalizable Higgs superpotential
has the form: 
\begin{equation}
W(\Phi ,\Psi )=M_{\Phi }\Phi^2+\lambda \Phi^3+
M_{\Psi }\overline {\Psi }^m\Psi_m+
h\Phi \overline {\Psi }^m\Psi_m    
\label {W84}
\end{equation}
We assume that the mass parameters $M_{\Phi }$, $M_{\Psi }$ are of
order of the GUT scale. Hence, the model contains only one mass 
scale $\sim M_G$ and no intermediate scale will arise in the model, 
unlike the model of ref. \cite{gia}. 

The VEV of $\Phi$ can induce the $SU(6)$ breaking only to the $G_{331}$
subgroup, among all maximal subgroups. In other words,  $SU(6)$ breaking to 
$SU(5)\times U(1)$ and $SU(4)\times SU(2)\times U(1)$ is not possible.
Indeed, decomposition of 175 in terms of the $G_{331}$ multiplets is:
\begin{eqnarray}
& 175=(1, 1)_{0}+(1, 1)_{-6}+(1, 1)_{6}+(8, 8)_0+
(6, \bar 6)_2+(\bar 6,6)_{-2}+   \nonumber \\ 
& +(3, \bar 3)_2+(\bar 3, 3)_{-2}+(3, \bar 3)_{-4}+(\bar 3, 3)_4  
\label{175}
\end{eqnarray}
where subscripts denote the $U(1)_I$ charges defined by the  
$Y_I\sim {\rm Diag}(1, 1, 1, -1, -1, -1)$ generator of $SU(6)$.  
It is crucial that 175 does not contain $SU(2)_W$ doublet, 
and so non of its fragments can participate in the breaking
$G_{331}\to G_{321}$. 
The component $(1,1)_0$ from 175 has the following VEV structure:
\begin{eqnarray}
\langle\Phi ^{abc}_{a'b'c'}\rangle 
=3V\epsilon ^{abc}\epsilon_{a'b'c'}~,~~~~
\langle\Phi ^{ijk}_{i'j'k'}\rangle
=-3V\epsilon ^{ijk}\epsilon_{i'j'k'}  \nonumber \\
\langle\Phi ^{iab}_{i'a'b'}\rangle
=-V\delta ^i_{i'}\epsilon ^{abc}\epsilon_{ca'b'}~,~~~~
\langle\Phi ^{ija}_{i'j'a'}\rangle
=V\delta ^a_{a'}\epsilon ^{ijk}\epsilon_{ki'j'}   
\label{vac1}
\end{eqnarray}
where $\epsilon$ is the $SU(3)$ invariant antisymmetric tensor, $a, b, ...$
and $i, j, ...$ denote $SU(3)_C$ and $SU(3)_W$ indices respectively.

$SU(3)_W\times U(1)_I$ is farther broken to $SU(2)_W\times U(1)_Y$
by the VEVs of $\overline {\Psi^m}+\Psi_m$ in $G_{321}$
singlet components.
Due to the $SU(2)_{cus}$ symmetry their VEVs can be placed only on the 
first pair. Then the doublet-antidoublet which come from 
$\overline {\Psi}^1+\Psi_1$ will be goldstones. As far the 
doublets from 
the $\overline {\Psi}^2+\Psi_2$, they remain massless because they are 
related to the genuine Goldstone doublets by custodial
$SU(2)_{cus}$ symmetry.
Below we present two examples of such $SU(6)\times SU(2)_{cus}$
models, with R chosen as 84 or 210.
\vspace {0.2cm}

${\bf R=84}$: 
In this case $\overline {\Psi}, \Psi$ are three index tensors
$\Psi^A_{BC}$, where $\Psi^A_{BC}=-\Psi^A_{CB}$ and $\Sigma \Psi^A_{AB}=0$.
The content of 84 with respect for $SU(5)$ subgroup is:
\begin{equation}
84=24_{-5}+45_{1}+5_{1}+10_{7}     \label {84}
\end{equation}
where subscripts denote the $U(1)$ charge of generator
$Y'={\rm Diag}(1, 1, 1, 1, 1, -5)$
of $SU(6)$. Therefore if $24_{-5}$ has a nonzero VEV in $G_{321}$
direction the $Y'$ generator is broken while ordinary hypercharge
$Y={\rm Diag}(2, 2, 2, -3, -3, 0)$ remains unbroken.

Analyzing the superpotential (\ref {W84}), one can see that there 
is an unique non-trivial supersymmetry conserving minima (with 
vanishing F and D terms) with the VEV configurations 
$\langle\Phi\rangle$ and $\langle\Psi\rangle$ that imply the  
$SU(6)$ symmetry breaking to $G_{321}$. 
More explicitly, $\langle\Phi\rangle$ has 
a form (\ref {vac1}) while $\langle\Psi\rangle$ is the following:
\footnote {Here the indices 4 and 5 stand for $SU(2)_W$, 
and $1, 2, 3$ for the $SU(3)_C$. 
Index 6 corresponds to the broken sixth degree of freedom  
of the $SU(6)$ gauge group.}.
\begin{equation}
\langle\Psi^A_{BC,m}\rangle
=U\left [\delta^6_C[2\delta^A_B-5(\delta^A_4\delta^4_B+
\delta^A_5\delta^5_B)]-\delta^6_B[B\to C]\right ]\delta _{1m}  
\label {vev2}
\end{equation}
(The VEV of $\overline {\Psi }$ is the same). 
The magnitudes of these VEVs are the following: 
\begin{equation}
V=\frac {M_{\Psi }}{10h}~,~~~~
U=\frac {3}{5h}\left (\frac {6}{5}M_{\Phi}M_{\Psi}\right )^{1/2}  \label
{sol1}
\end{equation}
Since ${\rm Tr}\langle\Phi^3\rangle=0$, $V$ and $U$ 
do not depend on the constant $\lambda$.

We see from (\ref {84}) that 84 contains two doublets which are 
compressed in the fragments 5 and 45. After the $SU(6)$ symmetry 
breaking to $G_{321}$ one combination of these doublets, which 
in terms of the $\Psi$ fragments reads as  
\begin{equation}
h_w^m=\frac {1}{\sqrt {21}}(-2\Psi^c_{cw}+
3\Psi^{w'}_{w'w})^m
\label {doub}
\end{equation}
is massless, while another combination 
\begin{equation}
H^m_w=\frac {1}{\sqrt 7}(\Psi^c_{cw}+2\Psi^{w'}_{w'w})^m 
\label {doub2}
\end{equation}
has the mass of order $M_{X}$ 
(here $c$ is the $SU(3)_C$ index and w, w' are the $SU(2)_W$ indices). 
The same applies to the conjugated states $\bar h$ and $\bar H$. 
As far as the pairs $(\bar h,h)_m$, $m=1,2$ 
are related by the $SU(2)_{cus}$ symmetry they are both massless. 
 First pair $(\bar h+h)^1$ is a genuine Goldstone mode 
which is eaten up by the corresponding gauge superfield of $SU(6)$, 
while the second one survives after $G_{X}$ breaking as a pseudogoldstone 
mode which can get $\sim M_W$ mass only after the supersymmetry breaking.

Note, that the coupling $\Phi\Psi\overline {\Psi}$ does not affect 
the structure  for the VEV $\langle\Phi\rangle$ and it maintains 
the pattern (\ref {vac1}). 
Although the term $h\Phi \bar {\Psi}^m\Psi_m$ in eq. (\ref {W84})
violates the extra global symmetries, from the "view" of 
doublet-antidoublet fragments in $\bar {\Psi} + \Psi$ the VEV 
$\langle \Phi\rangle$ is a singlet. After substituting the VEVs of 
$\Phi$ and $\Psi,\overline {\Psi}$ in the superpotential,  
due to the $SU(2)_{cus}$ symmetry no mixing terms emerge between 
the doublet (antidoublet) modes of $\Psi _{1} (\bar {\Psi ^{1}})$  and 
$\Psi_2 (\bar{\Psi^2})$, Since $\Phi$ itself does not include
the doublet modes. 
So one doublet-antidoublet is massless pseudogoldstone until SUSY
is unbroken. In this context this situation resembles the
pseudogoldstone picture \cite{ino, ber, bar} but the difference is that
we do not have the $SU(6)\times SU(6)$ global symmetry in the 
superpotential, but due to the structure of $\Phi$ and 
$SU(2)_{cus}$ symmetry the doublets can be rotated away from the Higgs 
superpotential. This can not be done for triplets from $\Psi $
because $\Phi $ itself contains the triplet fragments
and there occurs the mixing between triplets (antitriplets) from 175 
and 84. Without loosing of generality one can choose the basis
in which the mass matrix has the form:
\begin{eqnarray}
&\hspace{-6mm}3_{175}~~~~3_{84_1}~~~~3'_{84_1}~~~~3_{84_2}~~~3'_{84_2}&
\nonumber \\
\begin{array}{ccccc}
\bar 3_{175} \\
\bar 3_{\overline {84}_1} \\
\bar 3'_{\overline {84}_1} \\
\bar 3_{\overline {84}_2} \\
\bar 3'_{\overline {84}_2}
\end{array}&
\hspace{-6mm}\left(
\begin{array}{ccccc}
M_{\Phi }& \alpha U& \beta U& 0& 0 \\
\alpha U& M_{\Psi }& 0& 0& 0 \\
\beta U& 0& M_{\Psi }& 0& 0 \\
0& 0& 0& M_{\Psi }& 0 \\
0& 0& 0& 0& M_{\Psi } \end{array}
\right)&                                          \label {mat}
\end{eqnarray}
where
\begin{equation}
{\rm Det}\left(
\begin{array}{ccc}
M_{\Psi }& \alpha U& \beta U \\
\alpha U& M_{\Psi }& 0 \\
\beta U& 0& M_{\Psi } \end{array}
\right)=0                            \label {det}
\end{equation}
($\alpha $ and $\beta $ are some Clebsch factors which are not
important for us). 
Therefore, one eigenstate is massless and is identified as a Goldstone
mode.  Exact calculations show that all other states have masses
of order $M_{X}$. 

Fermion sector can be arranged in the same manner as in ref. \cite{gia}:
the quark-lepton masses are generated from the following 
Yukawa superpotential:
\begin{equation}
W_Y=g_d\bar 6^m~15\overline {\Psi^n}\epsilon_{mn}+
\frac {g_u}{M_0}15~15~\Psi_m\Psi_n\epsilon_{mn}            \label {fer}
\end{equation}
Second coupling can be obtained by  heavy particle exchange mechanism
\cite{fro}; After introducing two 20-plets 
($\equiv 20^m$) the renormalizable
couplings which are responsible for generation of 
$g_u$ term in (\ref {fer}) are
\begin{equation}
W^0_Y=g_115~20^m~\Psi_m+M_020^m~20^n~\epsilon_{mn}           \label {fer1}
\end{equation}
After integrating out the heavy $20^m$ states below the $M_0$ scale
we are left with the $g_u$ effective coupling.

\vspace {0.3cm}

${\bf R=210}$:
In this case $210\equiv \Psi^{AB}_{A'B'C'}$ is antisymmetric with respect
to the up and down indices and $\Sigma \Psi^{AB}_{AB'C'}=0$.
In terms of $SU(5)$ 210 reads as:
\begin{equation}
210=75+50+45+40   \label {210}
\end{equation}
where the $G_{321}$ singlet is contained in 75 and the MSSM doublet 
fragment in 45.  

The Higgs superpotential still has the same form (\ref {W84}), however 
now the coupling $h\Phi \bar {\Psi} \Psi $ implies three invariants:
\begin{equation}
h\Phi \bar {\Psi} \Psi \equiv \sum^3_{i=1} h_iI_i           \label {inv0}
\end{equation}
where
\begin{eqnarray}
I_1&=&\bar \Psi^{ABC}_{A'B'}\Phi^{A_1B_1C_1}_{ABC}\Psi^{A'B'}_{A_1B_1C_1}
\nonumber \\
I_2&=&\bar
\Psi^{ABC}_{A'B'}\Phi^{A_1B_1C_1}_{ABC}\Psi^{A'B'}_{A_1B_1C_1} \nonumber \\
I_3&=&\bar \Psi^{ABC}_{A'B'}\Phi^{A_1B_1C_1}_{ABC}\Psi^{A'B'}_{A_1B_1C_1}
\label {inv}
\end{eqnarray}

 The $G_{321}$ invariant VEV of $\Psi (\overline {\Psi })$ has the form:
\begin{eqnarray}
\Psi^{12}_{126}=\Psi^{13}_{136}&=&\Psi^{23}_{236}=~U   \nonumber\\
\Psi^{14}_{146}=\Psi^{24}_{246}=\Psi^{34}_{346}&=&
\Psi^{15}_{156}=\Psi^{25}_{256}=\Psi^{35}_{356}=-U   \nonumber\\
\Psi^{45}_{456}&=&3U                                     \label {vev3}
\end{eqnarray}
Supersymmetric minima allows to have nonvanishing $V$ and $U$
with the magnitudes:
\begin{equation}
V=\frac {9M_{\Psi }}{8(9h_1-h_2+h_3)},~~~~
U=\frac {9}{4(9h_1-h_2+h_3)} \left (\frac {M_{\Psi }
M_{\Phi}}{2}\right )^{1/2}                                 \label {sol2}
\end{equation}
and the  "philosophy" is the same as in the 84-plets case:
after the $SU(6)$ gauge symmetry breaking two 
pairs of the Higgs doublets
$\bar h^m+h_m$ remain massless.
One pair is absorbed by the appropriate gauge fields which
became superheavy and the second one survives. Therefore
in the effective low-energy theory we will have one pair
of massless Higgs doublet-antidoublet.

In this case the couplings relevant for the 
quark and lepton masses are the following:
\begin{equation}
W_Y=\frac {g_d}{M_1}\bar 6^m~15\Phi \overline {\Psi^n}\epsilon_{mn}+
\frac {g_u}{M_0}~15~15~\Psi_m~\Psi_n\epsilon_{mn}        \label {fer2}
\end{equation}
$g_u$ term can be generated in the same manner as
was discussed in the 84-plet's case, while the renormalizable 
Yukawa
superpotential which is responsible for generation of the $g_d$ term
has the form:  
\begin{equation} 
W'_Y=g'\bar 6^m~105~\bar{\Psi}^n\epsilon_{mn}+ 
g''15~\overline {105}\Phi +M_1\overline{105}~105            \label {fer3} 
\end{equation} 
Integrating out the
heavy $\overline {105}+105$ states, the first term of eq. (\ref
{fer2}) is obtained with $g_d\sim g'g''$.

As we see in this case the fermion sector 
requires more complicated multiplets because it is 
impossible to write renormalizable Yukawa couplings
for down quarks and leptons.
However if the mass of 105-plets $M_1$ is $10^{18}$ GeV
order, then after their decoupling the effective
Yukawa constants for third generation of down quark and lepton 
will have just needed magnitude --
$\frac {M_X}{M_1}\sim 10^{-2}$. 
More detailed study of fermion masses in our
model will be presented elsewhere.
\vspace{2mm}

Concluding, we have suggested supersymmetric $SU(6)$ theory in which 
the DT splitting occurs naturally. Although this mechanism is based on the 
custodial symmetry, the lightness of Higgs doublets  has the
different origin then in the model \cite{gia}. Crucial feature is that
the 175 not contain the Higgs doublets, and consequently 
there emerges no mixing between doublet components of the $SU(6)$ 
symmetry breaking scalars. This feature allows to achieve the one 
point unification of the gauge couplings at the scale 
$M_G\sim 10^{16}$ GeV. 
(Recall, that in the model \cite{gia} this mixing was rendering 
the scale $M_I$ the middle geometrical, $M_I\sim \sqrt {M_GM_W}$). 
Therefore, the `missing doublet' multiplet 175 of $SU(6)$ is 
very attractive for the model building.
Its properties admit Higgs doublets to be massless till SUSY
is unbroken. 
Besides 175, the higher dimensional selfconjugate representations 
of $SU(6)$ which do not contain doublet fragments 
are 3963 and 4116 \cite{sla}, which is clearly too much. 
As far the larger unitary groups $SU(6+N)$, one can make sure
that they have no selfconjugate "missing doublet"
multiplets which could be used for the symmetry braking. 
Therefore, the $SU(6)$ group appears to be the single group in 
which the presented mechanism can be realized.  

\vspace {3mm}

{\bf {Acknowledgement}}

I am grateful to Z. Berezhiani, J. Chkareuli, G. Dvali,
I. Gogoladze and A. Kobakhidze for useful discussions
and important comments.


\begin{thebibliography}{99}

\bibitem{lan}  P. Langacker and M. Luo, Phys. Rev. D44 (1991) 817; \\
U. Amaldi, W. de Boer and H. Furstenau, Phys. Lett. B260 (1991) 447; \\
J. Ellis, S. Kelley and D. Nanopoulos, Phys. Lett. B260 (1991) 131.

\bibitem{dim} S. Dimopoulos and F. Wilczek, in Erice Summer Lectures,
Plenum, New York, 1981; \\
H. Georgi, Phys. Lett. B108 (1982) 283; \\
B. Grinstein, Nucl. Phys. B206 (1982) 387; \\
A. Masiero, D.V. Nanopoulos, K. Tamvakis and T. Yanagida,
Phys. Lett. B115 (1982) 380; \\ 
J.L. Lopez and D.V. Nanopoulos, hep-ph/9508253.

\bibitem{dim1}  S. Dimopoulous and F. Wilczek, NSF-ITP-82-07 (unpublished); \\
M. Srednicki, Nucl. Phys. B202 (1982) 327; \\
K.S. Babu and S.M. Barr, Phys. Rev. D48 (1993) 5354;
Phys. Rev. D50 (1994) 3529.

\bibitem{ino}
K. Inoue, A. Kakuto and T. Takano, Progr. Theor. Phys. 75 (1986) 664; \\
A. Anselm and A. Johansen, Phys. Lett. B200 (1988) 331; \\
R. Barbieri, G. Dvali and A. Strumia, Nucl. Phys. B391 (1993) 487.

\bibitem{ber}  Z. Berezhiani and G. Dvali,
Sov. Phys. Lebedev Institute Reports 5 (1989) 55;  \\
R. Barbieri, G. Dvali and M. Moretti, Phys. Lett. B312 (1993) 137; \\
Z. Berezhiani, C. Csaki and L. Randall, Nucl. Phys. B444 (1995) 61.

\bibitem{gia}  G. Dvali, Phys. Lett. B324 (1994) 59.

\bibitem{And} G.V. Anderson et al., Phys. Rev. D49 (1994) 3660;\\
K.S. Babu and R.N. Mohapatra, Phys. Rev. Lett. 74 (1995) 2418; \\
K.S. Babu and S.M. Barr, {\it ibid}. 75 (1995) 2088; \\
Z. Berezhiani, Phys. Lett. B355 (1995) 178.

\bibitem{bar}  R. Barbieri, G. Dvali, A. Strumia, Z. Berezhiani, L.Hall,
Nucl.Phys. B432 (1994) 49; \\
Z. Berezhiani, Phys. Lett. B355 (1995) 481.

\bibitem{175}  I. Gogoladze, A. Kobakhidze and Z. Tavartkiladze,
Phys. Lett. B372 (1996) 246.

\bibitem{fro} C.D. Froggatt and H.B. Nielsen, Nucl. Phys. B147 (1979) 277; \\
Z.G. Berezhiani, Phys. Lett. B129 (1983) 99;  B150 (1985) 177; \\
S. Dimopoulos, Phys. Lett. B129 (1983) 417.

\bibitem{sla}  R. Slansky, Phys. Rep. 79 (1981) 3.

\end{thebibliography}
\end{document}